\documentclass[aps,prd,twocolumn,nofootinbib,longbibliography,10pt]{revtex4}
\usepackage{graphicx,amsmath,amssymb,url}
\usepackage[colorlinks=true, linkcolor=blue, citecolor=blue, urlcolor=blue]{hyperref}
\usepackage{BOONDOX-cal,BOONDOX-frak}
\begin{document}
    \title{Post-Newtonian analysis of the quantum signatures of gravity }
\author{Tuhin Chatterjee}
\email{tuhin.chatterjee26@gmail.com}
\affiliation{Department of Astrophysics and High Energy Physics, S. N. Bose National Centre for Basic Sciences, JD Block, Sector-III, Salt Lake City, Kolkata-700 106, India}
\author{Soham Sen}
\email{sensohomhary@gmail.com}
\affiliation{Department of Astrophysics and High Energy Physics, S. N. Bose National Centre for Basic Sciences, JD Block, Sector-III, Salt Lake City, Kolkata-700 106, India}
\author{Sunandan Gangopadhyay}
\email{sunandan.gangopadhyay@gmail.com}
\affiliation{Department of Astrophysics and High Energy Physics, S. N. Bose National Centre for Basic Sciences, JD Block, Sector-III, Salt Lake City, Kolkata-700 106, India}
\begin{abstract}
\noindent 
In a recent work \href{https://doi.org/10.1103/PRXQuantum.2.010325}{PRX QUANTUM 2 (2021) 010325}, a new way of investigating quantum gravity signatures using quantum information theoretic techniques, have been proposed. The primary result of this analysis revealed that non-Gaussianity can arise only through the consideration of a quantum model for the gravity part. Compared to classical gravity, only quantum gravity can result in non-quadratic operators in the Hamiltonian which leads to the non-Gaussian behavior. In our current analysis, we have considered a more realistic scenario taking into effect leading order post-Newtonian corrections in the analysis. We have stayed with the same model of a Bose-Einstein condensate placed inside a harmonic trap potential which indeed works as the detector of the non-Gaussianity generated due to quantum gravitational effects. Bose-Einstein condensates are experimentally well studied; apart from being a single quantum system, they include Feshbach resonances, which helps tuning the strength of the electromagnetic interactions which in principle can be set to zero. This is important since it can help distinguish quantum gravity from electromagnetic interactions without affecting gravitational interactions, and any non-Gaussianity can then be solely attributed to quantum gravity. We observe that the signal to noise ratio gets slightly damped due to the post-Newtonian effects taken under consideration.
\end{abstract}
\maketitle
\section{Introduction}
\noindent In recent times, the interest in finding phenomenological aspects of quantum gravity models has sky rocketed. We already know that three of the four fundamental forces in nature have a well defined quantum field theory. However, a renormalizable quantum field theory of gravity does not exist still now. Some of the primary contenders of a true quantum field theory of gravities are loop quantum gravity \cite{LQG1,LQG2}, non commutative geometry \cite{NCG1,NCG2}, and string theory \cite{String1,String2}. All of these theories have some standard problems and none of them qualify as a true quantum theory of gravity. A simpler way to check for quantum gravity signatures is to look for low-energy effects of quantum gravity in tabletop experimental set ups. Recently, works like \cite{QGEM1,QGEM2,QGEM3,QGEM4,QG1,QG2,
QG3,QG4,QG5,QG6,BECPRXQuantum} have tried to investigate quantum gravitational effects in relativistic quantum systems. We have in our works \cite{QG5,QG6} investigated the response of a Bose-Einstein condensate when it interacts with gravitational fluctuations and found out that a Bose-Einstein condensate based gravitational detector is highly efficient compared to other interferometer \cite{QG1,QG2} and resonant bar detector based \cite{SpontaneousPRD} proposed gravitational wave detectors. It has widely been argued that a tabletop experiment  is unable to detect quantum signatures of gravity \cite{Dyson}. In order to distinguish quantum gravity from classical gravity, one needs to probe the Planck mass scale \cite{ChristodoulouRovelli}. The Planck mass scale can be probed in a matter of very few years \cite{IsGravityQuantum,GravityinLab}. As a result of these experimental advancements, recent developments in the field of testing quantum signatures of gravity have gained a completely new momentum. One of the more interesting proposal for testing quantum gravity signatures in the laboratory is the famous Bose-Marletto-Vedral experiment \cite{QGEM1,QGEM2,QGEM3}. In \cite{BECPRXQuantum}, a substantially different approach was followed for testing quantum gravity with quantum information theoretic techniques. The gravitational potential was considered as a back-reaction of quantum matter which eventually resulted in the gravitational potential to be operator valued and they observed that non-Gaussianity was created in the quantum field of matter. In continuous variable quantum information science non-Gaussianity acts as a key resource and the quantum information is encoded as physical degrees of freedom in it. It was thoroughly analyzed in \cite{BECPRXQuantum} that classical gravity is unable to create non-Gaussianity in the quantum field of matter. A tabletop test of quantum gravity was proposed in \cite{BECPRXQuantum} where a $4\times 10^9$ atom BEC was considered to be kept inside of a single-well potential where the quantum gravity signal scales linearly with the time of interaction and quadratically with the system-mass. One of the fundamental benefits of using quantum gases is that using external optical or magnetic fields the electromagnetic interactions can be manipulated such that any external electro magnetic interruptions can be eliminated. In our analysis, we have extended the model proposed in \cite{BECPRXQuantum} by considering the back ground to have post-Newtonian corrections. The primary reason for considering such a background lies in the fact that any kind of tabletop experiments, the background geometry is not strictly flat, rather deformed due to the experiment being conducted on the surface of the Earth. For a more robust approach, we have started from the minimally coupled interaction Hamiltonian where the background gravitational perturbation couples to the energy momentum tensor and eventually obtained the operator valued interaction Hamiltonian by raising the system wave function to operator status. Raising the wave function to operator status leads to the raising of the Newtonian and post-Newtonian potential to operator status. We have then analytically obtained the analytical expression of the third and fourth order cumulants which eventually leads us to the signal to noise ration (SNR) of the model system. The manuscript is arranged as follows.

\noindent  In section (\ref{S2}), we have discussed the background of the model and obtained the analytical expression of the interaction part of the Hamiltonian in terms of the Newtonian potential in the presence of post-Newtonian corrections. In section (\ref{S3}), we obtain the explicit expression for the interaction Hamiltonian in terms of the raising a lowering operators of the phonons of a Bose-Einstein condensate. In section (\ref{S4}), we then proceed to calculate the fourth order cumulant for the system being in coherent and squeezed states and then we have obtained the signal to noise ratio or SNR for model. Finally, in section (\ref{S5}), we conclude our results.
\section{Background analysis}\label{S2}
\noindent We start by obtaining the analytical expression for the interaction Hamiltonian in the post-Newtonian set up.
The matter part of the action for the model system while considering the background as $g_{\mu\nu}=\eta_{\mu\nu}+h_{\mu\nu}$ ($h_{\mu\nu}$ being the small spacetime deformation), can be expressed as
\begin{equation}\label{1.1}
S_M=\int d^4x\sqrt{-g}\left(\mathcal{L}_M-\frac{1}{2}h_{\mu\nu}T^{\mu\nu}+\mathcal{O}(h^2)\right)
\end{equation}
with $\mathcal{L}_M$ gives the base part of the Lagrangian density and $T_{\mu\nu}$ gives the energy momentum tensor for the Bose-Einstein condensate. From the Lagrangian density, one can eventually obtain the Hamiltonian density and the interaction part of the Hamiltonian can be obtained simply as
\begin{equation}\label{1.2}
H_{int}=-\frac{1}{2}\int d^3\vec{r} ~h^{\mu\nu}(\vec{r})T_{\mu\nu}(\vec{r})~.
\end{equation}
In the post-Newtonian order, we shall go step by step to obtain the analytical expression of $g^{00}$.
Considering $\phi$ to be the Newtonian potential, one can express $g_{00}$ up to Newtonian order as $g_{00}=-1-2\phi$ where $\eta_{00}=-1$ and $h_{00}^{N}=-2\phi$. Where $N$ in the superscript of $h_{00}$ denotes Newtonian correction term. In our analysis, we shall extend our analysis to the post-Newtonian order. The line element in the post-Newtonian order \cite{Weinberg} without considering any cross terms read
\begin{equation}\label{1.3}
ds^2=-(1+2\phi+2\phi^2+2\psi)dt^2+(\delta_{ij}-2\phi\delta_{ij})dx^idx^j~.
\end{equation}
From the above expression, one can then reads of the $g_{00}$ component as $g_{00}=-1-2\phi-2\phi^2-2\psi$ where both $\phi^2$ and $\psi$ are post-Newtonian contributions and $h_{00}=h_{00}^{N.}-2\phi^2-2\psi$. From the above expression, we need to obtain the expression for $g^{00}$ using the condition $g_{\mu\nu}g^{\nu\xi}=\delta_\mu^{~\xi}$ which gives us $g_{00}g^{00}=1$. Now, for small $h_{\mu\nu}$, $g^{\mu\nu}\simeq\eta^{\mu\nu}-h^{\mu\nu}$ where the post-Newtonian order is considered to be absorbed in $h^{\mu\nu}$. Using $g_{00}$ from eq.(\ref{1.3}), we obtain
\begin{equation}\label{1.4}
\begin{split}
(-1-2\phi-2\phi^2-2\psi)(\eta^{00}-h^{00})&=1\\
1+2\phi+2\phi^2+2\psi+h^{00}+2\phi h^{00}&=1
\end{split}
\end{equation}
where we have dropped the terms $2\phi^2 h^{00}$ and $2\psi h^{00}$ as they will contribute beyond the Newtonian order. If we express $h^{00}=h^{00}_{N.}+h^{00}_{P.N.}$, then up to Newtonian order it is easy to observe that $h^{00}_{N.}=-2\phi$. Then $h^{00}=-2\phi+h^{00}_{P.N.}$ and substituting this back in eq.(\ref{1.4}), we can obtain the analytical expression of $h^{00}_{P.N.}$ as
\begin{equation}\label{1.5}
h^{00}_{P.N.}=2\phi^2-2\psi.
\end{equation}
Then $h^{00}$ is given by $h^{00}=-2\phi+2\phi^2-2\psi$ and $g^{00}=-1-2\phi+2\phi^2-2\psi$.
Now, the energy momentum tensor for dust gives simply $T_{\mu\nu}={\rho,0,0,0}.$ The interaction Hamiltonian in eq.(\ref{1.2}) then takes the form given by
\begin{equation}\label{1.6}
\begin{split}
H_{int}&=-\frac{1}{2}\int d^3\vec{r}~h^{\mu\nu}(\vec{r})T_{\mu\nu}(\vec{r})\\
&=-\frac{1}{2}\int d^3\vec{r}~h^{00}(\vec{r})T_{00}(\vec{r})+0\\
&=-\frac{1}{2}\int d^3\vec{r}~(-2\phi+2\phi^2-2\psi)\rho\\
\implies H_{int}&=\int d^3\vec{r}\left(\phi(\vec{r})-\phi^2(\vec{r})+\psi(\vec{r})\right)\rho(\vec{r}).
\end{split}
\end{equation}
The analytical expressions for the post-Newtonian correction term $\psi(\vec{r})$ takes the form \cite{Weinberg}
\begin{equation}\label{1.7}
\psi(\vec{r})=-G\int \frac{\left(T_{00}^{(2)}+T_{ii}^{(2)}\right)}{|\vec{r}-\vec{r}'|}d^3\vec{r}'
\end{equation}
where $(2)$ in the subscript of the stress-tensor simply denotes $\frac{v^2}{c^2}$ order terms. Setting $T_{00}^{(2)}=T_{ii}^{(2)}=0$, we can simply set $\psi(\vec{r})=0$ without loss of any generality. Raising $\phi(\vec{r})$ to operator status as well as $\rho(\vec{r})$, one can write down the interaction part of the Hamiltonian in operator representation as
\begin{equation}\label{1.8}
\hat{H}_{\text{int}}= \int d^3\vec{r}\left(\hat{\phi}(\vec{r})-\hat{\phi}^2(\vec{r})\right)\hat{\rho}(\vec{r})
\end{equation}
with the potential operator $\hat{\phi}(\vec{r})$ being denoted as
\begin{equation}\label{1.9}
\hat{\phi}(\vec{r})=-Gm_B\int d^3 \vec{r}'\frac{\hat{\Psi}^\dagger(\vec{r}')\hat{\Psi}(\vec{r}')}{|\vec{r}-\vec{r}'|}
\end{equation}
with $m_B$ denoting the mass of the system in consideration. With the interaction part of the Hamiltonian in hand, we now need to discuss the analytical expression for the wave function operator $\hat{\Psi}(\vec{r})$. 
\section{Quantum Gravity through Non-Gaussianity in Bose-Einstein-Condensate}\label{S3}
\noindent As has been discussed in \cite{BECPRXQuantum}, a Bose-Einstein condensate can always be described by a non-relativistic scalar field $\hat{\Psi}(\vec{r})$. A spherical Bose-Einstein condensate can now be considered with a finite radius $R$ where the system is allowed to interact with itself for a finite time. Now, if gravitational perturbations behave classically then it was argued in \cite{BECPRXQuantum} that the quantum state of the system will preserve its Gaussian behaviour, however, for a quantum gravitational set-up non-Gaussianity creeps in the quantum state of the system. In this analysis, we consider an identical set-up, however, as the system is in presence of the Earth's gravitational field it is more realistic to look for post-Newtonian contributions to the quantum state. We can now consider a Bose-Einstein condensate described by a scalar field $\hat{\Psi}(\vec{r})$ and it is possible to expand the scalar field as
\begin{equation}\label{1.10}
\hat{\Psi}(\vec{r})=\sum_i\Psi_i(\vec{r})\hat{a}_i=\Psi_0(\vec{r})\hat{a}_0+\sum_{i\neq 0}\Psi_i(\vec{r})\hat{a}_i~.
\end{equation}
Now the Bogoliubov approximation simply replaces the operator $\hat{a}_0$ by $\sqrt{N_0}$,
 where $\Psi_0(r)$ denotes the condensate wave function. As this model consists of an effective single particle, it is possible to neglect all other terms for $i\neq 0$ and one can approximate the field $\hat{\Psi}(\vec{r})$ in eq.(\ref{1.10}) simply by
 \begin{equation}\label{1.11}
\hat{\Psi}(\vec{r})\simeq \Psi_0(\vec{r})\hat{a}_0
 \end{equation}
where the annihilation operator $\hat{a}_0$ annihilates an atom at a position $\vec{r}$ \cite{BECPRXQuantum}. 

\noindent In a classical setting where the Newtonian gravitational potential behaves classically the interaction Hamiltonian simply takes the form
\begin{equation}\label{1.12}
\hat{H}_{\text{int}}^{\text{CG}}=\int d^3\vec{r} (\phi(\vec{r})-\phi^2(\vec{r}))\hat{\rho}(\vec{r})
\end{equation}
where only the density matrix remains operator valued $\hat{\rho}(\vec{r})=m_Bc^2\hat{\Psi}^\dagger(\vec{r})\hat{\Psi}(\vec{r})=m_Bc^2|\Psi_0(\vec{r})|^2\hat{a}^\dagger_0\hat{a}_0$. For the quantum gravitational case, however, the Newtonian gravitational potential becomes operator valued as can be seen from eq.(\ref{1.9}) and the normal ordered interaction Hamiltonian can be expressed as
\begin{equation}\label{1.13}
\hat{H}_{\text{int}}^{\text{QG}}= \int d^3\vec{r}:\hat{\phi}(\vec{r})\hat{\rho}(\vec{r}):-\int d^3\vec{r}:\hat{\phi}^2(\vec{r})\hat{\rho}(\vec{r}):~.
\end{equation}
For the post-Newtonian case, the classical interaction Hamiltonian takes the form
\begin{equation}\label{1.14}
\hat{H}_{\text{int}}^{\text{CG}}=\left(\lambda_{N}^{\text{CG}}[\Psi]+\lambda_{PN}^{\text{CG}}[\Psi]\right)\hat{a}_0^\dagger\hat{a}_0
\end{equation}
where the coefficients $\lambda_{N}^{\text{CG}}[\Psi]$ and $\lambda_{PN}^{\text{CG}}[\Psi]$ are given by
\begin{equation}\label{1.15}
\begin{split}
\lambda_{N}^{\text{CG}}[\Psi]=&m_B\int d^3\vec{r}|\Psi_0(\vec{r})|^2\phi[\Psi](t,\vec{r})\\
\lambda_{PN}^{\text{CG}}[\Psi]=&-m_B\int d^3\vec{r}|\Psi_0(\vec{r})|^2(\phi[\Psi](t,\vec{r}))^2
\end{split}
\end{equation}
where the speed of light is set to unity. In our current analysis, we are primary interested in the spacetime behaving like quantum entity due to the back reaction of the condensate with the background geometry and making use of eq.(s)(\ref{1.11},\ref{1.13}) the quantum gravitational interaction Hamiltonian reads
\begin{equation}\label{1.16}
\hat{H}_{\text{int}}^{\text{QG}}=\lambda_N \hat{a}_0^{\dagger^2}\hat{a}_0^2+\lambda_{PN}\hat{a}_0^{\dagger^3}\hat{a}_0^3
\end{equation}
with the Newtonian and post-Newtonian coefficients being defined as
\begin{align}
\lambda_N\equiv&-Gm_B^2\int d^3\vec{r} d^3\vec{r}'\frac{|\Psi_0(\vec{r})|^2|\Psi_0(\vec{r}')|^2}{\left|\vec{r}-\vec{r}'\right|}\label{1.17}\\
\lambda_{PN}\equiv&-G^2m_B^3\int d^3\vec{r} d^3\vec{r}'d^3\vec{r}''\frac{|\Psi_0(\vec{r})|^2|\Psi_0(\vec{r}')|^2|\Psi_0(\vec{r}'')|^2}{\left|\vec{r}-\vec{r}'\right|\left|\vec{r}-\vec{r}''\right|}\label{1.18}
\end{align}
where $c=1$. One thing to note here is that this Hamiltonian operator $\hat{H}^{QG}_{\text{int}}$ looks analogous to Kerr-type interaction in Quantum optics except the Post-Newtonian correction also adds to the Non-Gaussianity in the system. We now need to determine the analytical expressions for the Newtonian and post-Newtonian coefficients for a given ground state wave function $\Psi_0(\vec{r})$ of the Bose-Einstein condensate. The analytical expression for the wave function $\Psi_0(\vec{r})$ reads \cite{Pitaevski_Stingari}
\begin{equation}\label{1.19}
\Psi_0(\vec{r})=\left(\frac{m_B\omega}{\pi\hbar}\right)^\frac{3}{4}e^{-\frac{m_B\omega r^2}{2\hbar}}~.
\end{equation}
Defining the characteristic length as $R\equiv\sqrt{\frac{\hbar}{m_B\omega}}$, we can express $\Psi_0(r)$ as $\Psi_0(\vec{r})=\Psi_0(r)=\frac{1}{\pi^\frac{3}{4}R^\frac{3}{2}}e^{-\frac{r^2}{2R^2}}$. At first, we need to evaluate the Newtonian coefficient and for simplicity we consider that the relative angle $\theta$ is not correlated to $\theta'$. Hence, it is possible to run both the angular integrals from $0$ to $\pi$.
\footnote{One can also proceed without this assumption which just slightly modifies the Newtonian coefficient in the interaction Hamiltonian by a numerical factor.} We then arrive at the analytical expression for $\lambda_N$ from eq.(\ref{1.17}) as
\begin{widetext}
\begin{equation}\label{1.20}
\lambda_N=-\frac{(2\pi)^2Gm_B^2}{\pi^3R^6}\int_0^{\infty}dr r^2 e^{-\frac{r^2}{R^2}}\int_0^\pi d\theta~\sin\theta \int_0^rdr' {r'}^2e^{-\frac{{r'}^2}{R^2}}\int_0^{\pi} d\theta' \frac{\sin\theta'}{\sqrt{r^2+{r'}^2-2rr'\cos\theta'}}
\end{equation}
\end{widetext}
where the azimuthal integrals have been executed. For the $\theta'$ integral, we obtain
\begin{equation}\label{1.21}
\begin{split}
I_{\theta}(r,r')=&\int_0^\pi d\theta' \frac{\sin\theta'}{\sqrt{r^2+{r'}^2-2rr'\cos\theta'}}\\=&\frac{\sqrt{(r+r')^2}-\sqrt{(r-r')^2}}{rr'}
\end{split}~.
\end{equation}
For $r>r'$, $I_\theta=\frac{2}{r}$ and for $r<r'$, $I_\theta=\frac{2}{r'}$. Because of the Gaussian decay factor the upper limits of radial integrals can safely be extended to infinity giving the overall analytical expression for $\lambda_N$ to be
\begin{equation}\label{1.22}
\lambda_N=-\sqrt{\frac{2}{\pi}}\frac{Gm_B^2}{R}~.
\end{equation}
Similarly, one can obtain the analytical expression for the post-Newtonian coefficient $\lambda_{PN}$ as
\begin{equation}\label{1.23}
\lambda_{PN}=-\frac{2G^2m_B^3}{3R^2}~.
\end{equation}
We now define two new parameters $\chi_N$ and $\chi_{PN}$ as
\begin{align}
\chi_N=&\frac{|\lambda_N|}{\hbar}=\sqrt{\frac{2}{\pi}}\frac{Gm_B^2}{\hbar R},~\text{and}~
\chi_{PN}=\frac{|\lambda_{PN}|}{\hbar}=\frac{2G^2m_B^3}{3\hbar R^2}\label{1.24}.
\end{align}
It is important to note that both $\chi_N$ and $\chi_{PN}$ have the dimension of frequency and for a condensate with fundamental phonon frequency $\omega\sim 10^2$ Hz and $m_B\sim 10^{-25}$ kg, one can obtain the numerical values of the Newtonian and Post-Newtonian corrections to the base frequency as $\chi_N\sim 10^{-22}$ Hz and $\chi_{PN}\sim 10^{-52}$ Hz. It is therefore evident that $\chi_{N}\gg\chi_{PN}$. If now, a condensate with $N_0\sim 10^9$ number of particles is considered then $\chi_N\rightarrow N_0\chi_N\sim 10^{-13}$ Hz and the post-Newtonian correction frequency goes as $\chi_{PN}\rightarrow N_0^2\chi_{PN}\sim 10^{-34}$ Hz. In the next section, we shall discuss in details the observational strategy for such a model \footnote{Here we consider a very weakly interacting BEC (coupling parameter $g\simeq0$) with the individual bosonic mass being $m_B$ and the harmonic trap frequency being $\omega$.}. 
\section{Observation strategy}\label{S4}
\noindent We start with the interaction Hamiltonian for the classical case in eq.(\ref{1.14}). The important thing to remember is that for the interaction part of the Hamiltonian the total Hamiltonian get re-modified where the base frequency of the model system gets renormalized and the Gaussian behavior remains intact. The primary being the interaction Hamiltonian remains proportional to the number operator. For the quantum gravitational interaction Hamiltonian, however, the interaction is no longer just proportional to the number operator which results in the non-Gaussian behavior. It is thus evident that signatures of non-Gaussianity will provide a direct evidence for the existence of the quantum nature of gravity. The aim is therefore to probe this non-Gaussianity for experimental verification. For a purely Gaussian distribution one can purely characterize it by calculating the first and second order ($\kappa_1$ and $\kappa_2$) cumulants and any higher order cumulants vanish ($\kappa_n=0$ $\forall n>2$). If the higher order cumulants are found to be non-zero it directly indicates towards the skewness of the Gaussian distribution leading to signature of quantum gravity. For our Bose-Einstein condensate based system, the generalized quadrature can be defined as $\hat{q}(\theta)=\frac{1}{\sqrt{2}}\left(\hat{a}_0e^{-i\theta}+\hat{a}_0^\dagger e^{i\theta}\right)$. We can start with third order cumulant, however, it is more prudent to calculate the fourth order cumulant as the third order cumulant $\kappa_3$ can be zero for symmetric non-Gaussian distributions and thus it may not capture the non-Gaussianity when the time evolution is symmetric \cite{BECPRXQuantum}. The fourth order cumulant in terms of the quadrature $\hat{q}$ is given by \cite{k_Statistics}
\begin{equation}\label{1.25}
\kappa_4 =\langle \hat{q}^4\rangle-4\langle\hat{q}^3\rangle\langle \hat{q}\rangle-3\langle\hat{q}^2\rangle^2+12\langle\hat{q}^2\rangle\langle\hat{q}\rangle^2-6\langle \hat{q}\rangle^4
\end{equation}
For an experimental scenario, one can use a finite sample from which one needs to estimate $\kappa_4$. For a choice of unbiased estimators, one needs to use the $k$-statistics such that $\kappa_n=\langle k_n\rangle$. For our model it is more logical to use coherent state for taking the expectation value as the condensate is primarily in a coherent state. In our model however, we have also considered squeezed state and the results are given in an Appendix. For the given measurement and experimental set-up, one can define the signal to noise ratio or SNR as \cite{BECPRXQuantum}
\begin{equation}\label{1.26}
    \text{SNR}\equiv\frac{|\kappa_4|}{\sqrt{\text{var} [k_4]}}
\end{equation}
where we wish to maximize the SNR for better accuracies. The analytical expression for the variance of $k_4$ can be obtained to be\cite{k_Statistics,BECPRXQuantum}
\begin{equation}\label{1.27}
\begin{split}
&\text{var}[k_4]\\&= \frac{\kappa_8}{\mathcal{M}}+\frac{16\kappa_2\kappa_6}{\mathcal{M}-1}+\frac{48\kappa_3\kappa_5}{\mathcal{M}-1}+\frac{34\kappa_4^2}{\mathcal{M}-1}+\frac{72\kappa_2^2\kappa_4\mathcal{M}}{(\mathcal{M}-1)(\mathcal{M}-2)}\\&+\frac{144\kappa_2\kappa_3^2\mathcal{M}}{(\mathcal{M}-1)(\mathcal{M}-2)}+\frac{24\kappa_2^4\mathcal{M}(\mathcal{M}+1)}{(\mathcal{M}-1)(\mathcal{M}-2)(\mathcal{M}-3)}
\end{split}
\end{equation}
where $\mathcal{M}$ is the number of independent estimation and we have considered $\mathcal{M}\gg 1$ for our current model. In eq.(\ref{1.27}), $k_4$ denotes the fourth statistics and the $n$-th order cumulant $\kappa_n$ can be obtained as
\begin{align}\label{1.28}
    \kappa_n=\langle\hat{q}^n\rangle-\sum_{m=1}^{n-1}\binom{n-1}{m-1}\langle\hat{q}^{n-m}\rangle\kappa_m~.
\end{align}
It is important to note that for our model the leading base line remains Gaussian and hence in the denominator only the leading Gaussian contribution actually contributes to the signal to noise ratio. For a Gaussian profile $\kappa_n=0$ if $n>2$. Hence, effectively the variance expression in eq.(\ref{1.27}) reduces to
\begin{equation}\label{1.29}
\text{var}[k_4]\simeq \frac{24 \kappa_2^4}{\mathcal{M}}
\end{equation}
where $\mathcal{M}\gg 1$ condition is implemented. For coherent state ($\hat{a}_0|\alpha\rangle=\alpha|\alpha\rangle$), $\kappa_2=\langle\hat{q}^2\rangle-\langle\hat{q}\rangle^2=\frac{1}{2}$. Substituting $\kappa_2$ in eq.(\ref{1.29}), we obtain the analytical expression for the leading order term of the denominator in SNR expression in eq.(\ref{1.26}) as $\sqrt{\text{var}[k_4]}=\sqrt{\frac{3}{2\mathcal{M}}}~.$
We are now in a position to proceed to calculate the fourth order cumulant for the system being initially in a coherent state.
\subsection{Calculations for the Fourth-order-cumulant}
\subsubsection{Coherent state}
\noindent The analytical expression for the fourth order cumulant is given in eq.(\ref{1.25}) and the expression reads
$\kappa_4=\langle \hat{q}^4\rangle-4\langle\hat{q}^3\rangle\langle \hat{q}\rangle-3\langle\hat{q}^2\rangle^2+12\langle\hat{q}^2\rangle\langle\hat{q}\rangle^2-6\langle \hat{q}\rangle^4$ where the quadrature reads $\hat{q}=\frac{1}{\sqrt{2}}\left(\hat{a}e^{-i\theta}+\hat{a}^\dagger e^{i\theta}\right)$. We now need to analytically obtain the time evolution of the ladder operators $\hat{a}_0$ and $\hat{a}_0^\dagger$. The total quantum gravitational Hamiltonian for a single mode Bose-Einstein condensate without any electromagnetic interactions read
\begin{equation}\label{1.30}
\begin{split}
:\hat{H}:=&:[\hat{H}_0+\hat{H}^{\text{QG}}_{\text{int}}]:=\hbar \omega \hat{a}_0^\dagger\hat{a}_0+\lambda_N \hat{a}_0^{\dagger^2}\hat{a}_0^2+\lambda_{PN}\hat{a}_0^{\dagger^3}\hat{a}_0^3~.
\end{split}
\end{equation}
As we are considering condensate system, we assume that the total number of particles is very large and constant in time and as a result the number operator $\hat{N}(t)=\hat{a}_0^\dagger(t)\hat{a}_0(t)=\hat{a}_0^\dagger\hat{a}_0=\hat{N}$ remains time independent.
For example, under the action of the above Hamiltonian, the time evolution equation for the operator $\hat{a}_0^m(t)$ takes the analytical form
\begin{equation}\label{1.31}
    \frac{d\hat{a}_0^n}{dt}=-\frac{i}{\hbar}[\hat{a}_0^n,\hat{H}]~.
\end{equation}
Solving the above equation, one can write down the analytical expression for $\hat{a}^n_0(t)$ in terms of the ladder operator at time $t=0$ as
\begin{widetext}
\begin{equation}\label{1.32}
\hat{a}^n_0(t)=\exp\left[-i (\omega n-\chi_Nn(n-1)-\chi_{PN}n (n-1)(n-2))t+i\left(2\chi_Nn+3\chi_{PN}n(n-2)\right)\hat{N}t+3i\chi_{PN}n\hat{N}^2t\right]\hat{a}^n_0
\end{equation}
where $\hat{a}^n_0=\hat{a}^n_0(0)$. Similarly the $n$-th order time dependent raising operator takes the analytical expression
\begin{equation}\label{1.33}
\hat{a}^{\dagger^n}_0(t)=\hat{a}^{\dagger^n}_0\exp\left[i (\omega n-\chi_N n(n-1)-\chi_{PN}n (n-1)(n-2))t-i\left(2\chi_Nn+3\chi_{PN}n(n-2)\right)\hat{N}t-3i\chi_{PN}n\hat{N}^2t\right]
\end{equation}
where $\hat{a}^{\dagger^n}_0=\hat{a}^{\dagger^n}_0(0)$. 

\noindent Finally, we can write down the expression for the mixed operators $\hat{a}^{\dagger^m}_0(t)\hat{a}^n_0(t)$ as
\begin{equation}\label{1.34}
\begin{split}
\hat{a}^{\dagger^m}_0(t)\hat{a}^n_0(t)=&\hat{a}^{\dagger^m}_0e^{i (\omega (m-n)-\chi_N (m(m-1)-n(n-1))-\chi_{PN}(m (m-1)(m-2)-n(n-1)(n-2)))t}\\
&\times~e^{-i\left(2\chi_N(m-n)+3\chi_{PN}(m(m-2)-n(n-2))\right)\hat{N}t-3i\chi_{PN}(m-n)\hat{N}^2t}\hat{a}^n_0~.
\end{split}
\end{equation}
\end{widetext}
It is evident from the above expression in eq.(\ref{1.34}) that for $m=n$, $\hat{a}^{\dagger^n}_0(t)\hat{a}^n_0(t)=\hat{a}^{\dagger^n}_0\hat{a}^n_0$.
We will be needing the expressions from eq.(\ref{1.32},\ref{1.33},\ref{1.34}) while evaluating the quadrature expectation values.

\noindent It is evident while calculating $\langle \hat{q}^n(t)\rangle$ we need to evaluate terms like $\langle\alpha|e^{-\beta_{\text{N}}\hat{N}-\beta_{\text{PN}}\hat{N}^2}|\alpha\rangle$. One can easily understand if we can express $e^{-\beta_{\text{N}}\hat{N}-\beta_{\text{PN}}\hat{N}^2}$ in a normalized form, it would be possible to easily take the expectation value with respect to the coherent states. It is important to note that $\beta_{\text{N}}$ may contain both $\chi_N$ and $\chi_{PN}$. For a generalized approach, we start with a general function of the number operator $\hat{N}$. Using standard analytical and algebraic approaches, we shall eventually try to obtain a normalized operator expectation which will let us replace any number operator just by $|\alpha|^2$.
\subsubsection{Derivation of the normal ordered exponential expression}
\noindent Any function of the number operator $\hat{N}$ can be expressed as
\begin{equation}\label{A.1}
f(\hat{N})=\sum_{l=0}^\infty \frac{f^{(l)}(0)}{l!}\hat{N}^l
\end{equation}
where $f^{(l)}(0)$ denotes the $l$-th order derivative of the function at the saddle point. It is possible to again express the normal ordering operator and its higher powers in a compact summation notation as
\begin{equation}\label{A.2}
\hat{N}^m=\sum_{j=0}^m S^{(j)}_m\hat{a}^{\dagger^j}\hat{a}^j
\end{equation}
where the analytical expression for $S_{m}^{(j)}$ is given by
\begin{equation}\label{A.3}
S_{m}^{(j)}=\frac{1}{j!}\sum_{k=0}^j(-1)^{j-k}\frac{j!}{k!(j-k)!}k^m
\end{equation}
and $S_{m}^{(j)}=0$ $\forall j>m$.
We substitute back the expression in eq.(\ref{A.3}) back in eq.(\ref{A.1}) as
\begin{equation}\label{A.4}
f(\hat{N})=\sum_{l=0}^\infty \frac{f^{(l)}(0)}{l!}\sum_{k=0}^l S_{l}^{(k)}\hat{a}^{\dagger^k}\hat{a}^k~.
\end{equation}
We already know that $\sum_{k=l+1}^\infty S_{l}^{(k)}\hat{a}^{\dagger^k}\hat{a}^k=0$ as $S_{l}^{(k)}$ vanishes for $k>l$. We can the rewrite eq.(\ref{A.4}) as
\begin{equation}\label{A.5}
\begin{split}
f(\hat{N})=&\sum_{l=0}^\infty \frac{f^{(l)}(0)}{l!}\sum_{k=0}^\infty S_{l}^{(k)}\hat{a}^{\dagger^k}\hat{a}^k\\
=&\sum_{k=0}^\infty\hat{a}^{\dagger^k}\hat{a}^k\sum_{l=k}^\infty \frac{f^{(l)}(0)}{l!}S_{l}^{(k)}
\end{split}
\end{equation}
where in the last line of the above expression, one needs to reuse the fact that $S_l^{(k)}=0$ $\forall k>l$.
Again substituting the expression for $S_{l}^{(k)}$ from eq.(\ref{A.3}), we can express the function $f(\hat{N})$ as
\begin{equation}\label{A.6}
f(\hat{N})=\sum_{k=0}^\infty\hat{a}^{\dagger^k}\hat{a}^k\sum_{l=k}^\infty \frac{f^{(l)}(0)}{l!k!}\sum_{p=0}^k (-1)^{k-p}\frac{k!}{p!(k-p)!}k^l~.
\end{equation}
Again $\Delta^p f(x)$ is given as \cite{GradshteynRyzik}
\begin{equation}\label{A.7}
\Delta^p f(x)=\sum_{j=0}^p(-1)^{p-j}\frac{p!}{j!(p-j)!}f(x+j)~.
\end{equation}
This helps us to write the function eq.(\ref{A.6}) as
\begin{equation}\label{A.8}
f(\hat{N})=\sum_{k=0}^\infty\hat{a}^{\dagger^k}\hat{a}^k\frac{\Delta^k f(0)}{k!}~.
\end{equation}
Consider $f(\hat{N})=e^{-\beta_\text{N}\hat{N}-\beta_\text{PN}\hat{N}^2}$. Using the functional form, we can then recast eq.(\ref{A.8}) as
\begin{equation}\label{A.9}
\Delta^k f(0)=\sum_{j=0}^k(-1)^{k-j}\frac{k!}{j!(k-j)!}e^{-\beta_\text{N}j-\beta_\text{PN}j^2}~.
\end{equation}
We now decompose the final exponential term with the post-Newtonian coefficient $\beta_{\text{PN}}$ as
\begin{equation}\label{A.10}
e^{-\beta_\text{PN}j^2}=\sum_{n=0}^\infty\frac{(-\beta_{\text{PN}})^n}{n!}j^{2n}~.
\end{equation}
Even if $\beta_{\text{PN}}$ is not small then also the above decomposition is valid as the radius of convergence for an exponential function is infinity, however, in our case $\beta_{\text{PN}}\ll 1$ which allows for truncation of the series up to a particular order of the post-Newtonian coefficient. Using the decomposition of the exponential term in eq.(\ref{A.10}) and substituting it in eq.(\ref{A.9}) and eventually replacing the analytical form of $\Delta^k f(0)$ in eq.(\ref{A.8}), we can write down the analytical expression for $f(\hat{N})$ as
\begin{widetext}
\begin{equation}\label{A.11}
f(\hat{N})=\sum_{k=0}^\infty\frac{\hat{a}^{\dagger^k}\hat{a}^k}{k!}\sum_{j=0}^k(-1)^{k-j}\frac{k!}{j!(k-j)!}e^{-\beta_{\text{N}}j}\sum_{n=0}^\infty\frac{(-\beta_{\text{PN}})^n}{n!}j^{2n}~.
\end{equation}
Now it is quite easy to check that $\frac{\partial e^{-\beta_{\text{N}}j}}{\partial \beta_{\text{N}}}=-je^{-\beta_{\text{N}}j}$. Hence, one can express $j^{2n}e^{-\beta_{\text{N}}j}$ as
\begin{equation}\label{A.12}
\begin{split}
\left(\frac{\partial}{\partial \beta_{\text{N}}}\right)^{2n}e^{-\beta_{\text{N}}j}=&(-j)^{2n}e^{-\beta_{\text{N}}j}\\=&j^{2n}e^{-\beta_{\text{N}}j}
\end{split}
\end{equation}
where $(-1)^{2n}=1$ as $n$ is an integer. Using the above expression, we can recast the expression in eq.(\ref{A.11}) as
\begin{equation}\label{A.13}
\begin{split}
f(\hat{N})=&\sum_{k=0}^\infty\frac{\hat{a}^{\dagger^k}\hat{a}^k}{k!}\sum_{n=0}^\infty\frac{(-\beta_{\text{PN}})^n}{n!}\left(\frac{\partial}{\partial \beta_{\text{N}}}\right)^{2n}\left[\sum_{j=0}^k(-1)^{k-j}\frac{k!}{j!(k-j)!}e^{-\beta_{\text{N}}j}\right]\\
=&\sum_{n=0}^\infty\frac{(-\beta_{\text{PN}})^n}{n!}\left(\frac{\partial}{\partial \beta_{\text{N}}}\right)^{2n}\left[\sum_{k=0}^\infty\frac{\hat{a}^{\dagger^k}\hat{a}^k}{k!} (e^{-\beta_{\text{N}}}-1)^k\right]
\end{split}
\end{equation}
\end{widetext}
where in the last line of the above expression, we have used the identity 
\begin{equation}\label{A.14}
\sum_{j=0}^k(-1)^{k-j}\frac{k!}{j!(k-j)!}e^{-\beta_{\text{N}}j}=(e^{-\beta_{\text{N}}}-1)^k~.
\end{equation}
We can further simplify the expression in eq.(\ref{A.13}), and write it in a more compact notation as
\begin{equation}\label{A.15}
\begin{split}
f(\hat{N})=e^{-\beta_{\text{PN}}\left(\frac{\partial}{\partial{\beta_{\text{N}}}}\right)^2}:e^{(e^{-\beta_{\text{N}}}-1)\hat{N}}:~.
\end{split}
\end{equation}
The above expression denotes the compact expression for the function $f(\hat{N})$ in a normal ordered notation.
As the exponential term outside of the normal ordering does not contain any operatorial contribution, we can put it back inside of the normal ordering to express the entire expression in a compact notation as
\begin{equation}\label{A.16}
e^{-\beta_\text{N}\hat{N}-\beta_{\text{PN}}\hat{N}^2}=:e^{-\beta_{\text{PN}}\frac{\partial^2}{\partial\beta_{\text{N}}^2}}e^{\left(e^{-\beta_{\text{N}}}-1\right)\hat{N}}:~.
\end{equation}
In the limit $\beta_{\text{PN}}\rightarrow 0$, we get back the usual normal ordering result for the Newtonian case \cite{BECPRXQuantum} which is $e^{-\beta_\text{N}\hat{N}}=:e^{\left(e^{-\beta_{\text{N}}}-1\right)\hat{N}}:$.
The derived normalized form of the operator in eq.(A.16) is one of the most important expressions in our manuscript and is central to our fourth cumulant calculation. Before proceeding to obtain the final expression for $\kappa_4$, we need to remember that $\alpha=\sqrt{N_0}e^{-i\theta_0}$ and it is simply possible to set the phase of the eigenvalue to zero such that $\alpha=\alpha^*=\sqrt{N_0}$. We will now list the expectation values $\langle\hat{q}^n\rangle$ up to $n=4$ and eventually write down the final simplified expression for $\kappa_4$. 
\begin{widetext}
The expectation value of $\hat{q}$ is obtained as
\begin{equation}\label{1.51}
\langle \hat{q}\rangle=\sqrt{\frac{N_0}{2}}e^{-i(\theta+\omega t)}e^{-\mathcal{B}_1\frac{\partial^2}{\partial\mathcal{A}_1^2}}e^{\left(e^{-\mathcal{A}_1}-1\right)N_0}+\sqrt{\frac{N_0}{2}}e^{i(\theta+\omega t)}e^{-\mathcal{B}^*_1\frac{\partial^2}{\partial\mathcal{A}^{*^2}_1}}e^{\left(e^{-\mathcal{A}^*_1}-1\right)N_0}
\end{equation}
where $\mathcal{A}_1\equiv-2i\chi_Nt+3i\chi_{PN}t$ and $\mathcal{B}_1\equiv-3i\chi_{PN}t$. Similarly the other expectation values read
\begin{align}
\langle \hat{q}^2\rangle=&\frac{N_0}{2}e^{-2i(\theta+(\omega-\chi_N) t)}e^{-\mathcal{B}_2\frac{\partial^2}{\partial\mathcal{A}_2^2}}e^{\left(e^{-\mathcal{A}_2}-1\right)N_0}+\frac{N_0}{2}e^{2i(\theta+(\omega-\chi_N) t)}e^{-\mathcal{B}^*_2\frac{\partial^2}{\partial\mathcal{A}^{*^2}_2}}e^{\left(e^{-\mathcal{A}^*_2}-1\right)N_0}+N+\frac{1}{2}\label{1.52}\\
\langle\hat{q}^3\rangle=&\left[\frac{N_0}{2}\right]^\frac{3}{2}\left[6+e^{-3i(\theta+(\omega-2\chi_N-2\chi_{PN}) t)}e^{-\mathcal{B}_3\frac{\partial^2}{\partial\mathcal{A}_3^2}}e^{\left(e^{-\mathcal{A}_3}-1\right)N_0}+3e^{-i(\theta+(\omega-2\chi_N) t)}e^{-\mathcal{D}_3\frac{\partial^2}{\partial\mathcal{C}_3^2}}e^{\left(e^{-\mathcal{C}_3}-1\right)N_0}+\text{H.C.}\right]+\frac{3}{2}\langle\hat{q}\rangle\label{1.53}\\
\langle\hat{q}^4\rangle=&\frac{N_0^2}{4}\left[e^{-4i(\theta+(\omega-3\chi_N-6\chi_{PN}) t)}e^{-\mathcal{B}_4\frac{\partial^2}{\partial\mathcal{A}_4^2}}e^{\left(e^{-\mathcal{A}_4}-1\right)N_0}+4e^{-2i(\theta+(\omega-3\chi_N-3\chi_{PN}) t)}e^{-\mathcal{D}_4\frac{\partial^2}{\partial\mathcal{C}_4^2}}e^{\left(e^{-\mathcal{C}_4}-1\right)N_0}+\text{H.C.}\right]\nonumber\\
+&\frac{1}{4}\left[6N_0e^{-2i(\theta+(\omega-\chi_N) t)}e^{-\mathcal{B}_2\frac{\partial^2}{\partial\mathcal{A}_2^2}}e^{\left(e^{-\mathcal{A}_2}-1\right)N_0}+\text{H.C.}+12N_0^2+3\right]
\end{align}
where the coefficients are defined as $\mathcal{A}_2\equiv-4i\chi_Nt$, $\mathcal{B}_2\equiv-6i\chi_{PN}t$, $\mathcal{A}_3\equiv-6i\chi_Nt-9i\chi_{PN}t$, $\mathcal{B}_3\equiv-9i\chi_{PN}t$, $\mathcal{C}_3\equiv-2i\chi_Nt-3i\chi_{PN}t$, $\mathcal{D}_3\equiv-3i\chi_{PN}t$, $\mathcal{A}_4\equiv-8i\chi_Nt-24i\chi_{PN}t$, $\mathcal{B}_4\equiv-12i\chi_{PN}t$, $\mathcal{C}_4\equiv-4i\chi_Nt-12i\chi_{PN}t$, and $\mathcal{D}_4\equiv-6i\chi_{PN}t$.
\end{widetext}
We can now set the phase $\theta$ to $\frac{\pi}{2}$ and $\omega$ to be very small. The reason for setting $\omega$ to be very small lies in the fact that the base frequency only contributes to the Gaussian profile and has not direct contribution towards skewness. Keeping terms up to $\mathcal{O}(\chi_{PN})$, we can then obtain the leading order analytical expression for the fourth cumulant $\kappa_4$ as
\begin{equation}\label{1.55}
\kappa_4\simeq 192 N_0^3 \chi_N^3 t^4(9\chi_N+197N_0\chi_{PN})-72N_0^2 \chi_N\chi_{PN}t^2~.
\end{equation}
The above expression is indeed very interesting as it may seem that the leading order contribution comes from the $72N_0^2 \chi_N\chi_{PN}t^2$ term which contains a post-Newtonian correction frequency as $\chi N_0 t\ll 1$. Our analysis, however, ensures that $\chi_{N}\gg\chi_{PN}$ which holds true if and only if the measurement time is above a certain cut-off value $t_{\text{min}}$. It is important to note that at $t=t_{\text{min}}$ $\kappa_4=0$ which indicates a zero value for the SNR. The SNR expression using eq.(s)(\ref{1.26},\ref{1.29},\ref{1.55}) can be then obtained as
\begin{equation}\label{1.56}
\text{SNR}\simeq \sqrt{\frac{2\mathcal{M}}{3}}\left|1728N_0^3\chi_N^4 t^4-72N_0^2\chi_N\chi_{PN}t^2\right|~.
\end{equation}
The most important thing to understand is that the SNR is always positive and for $t<t_{\text{min}}$ the post-Newtonian contribution should theoretically dominate, however, the total SNR becomes very small keeping it in the undetectable zone.
The minimum value of time can then be obtained as $t_{\text{min}}=\frac{1}{2}\sqrt{\frac{\chi_{PN}}{6N_0\chi_N^3}}\sim 1.71\times 10^{-3}$ sec.
\begin{figure}[ht!]
\begin{center}
\includegraphics[scale=0.35]{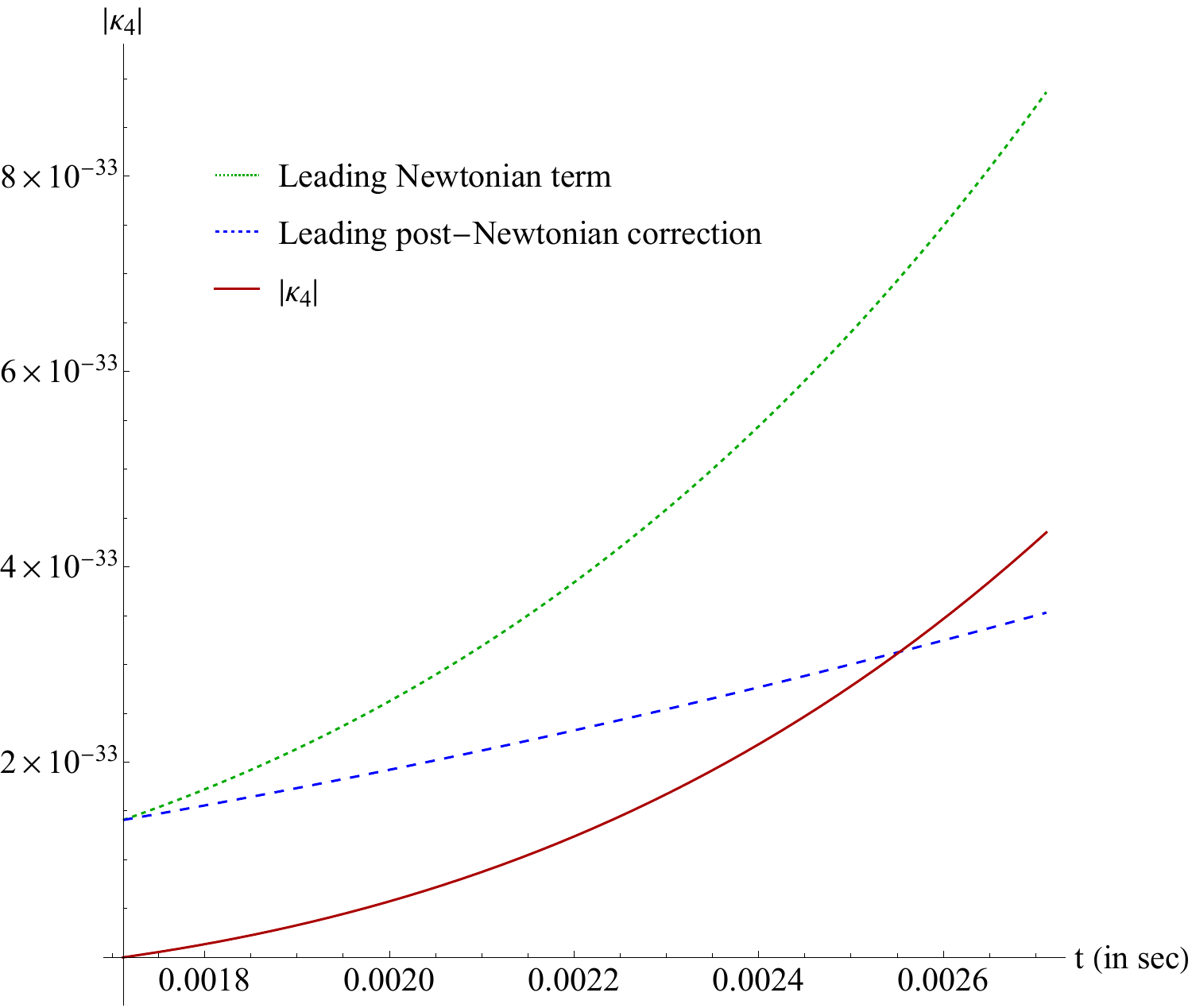}
\caption{Plot of the fourth order cumulant against the total time of observation and comparison with the leading order Newtonian and post-Newtonian contributions.\label{PN}}
\end{center}
\end{figure}
We now plot the leading order Newtonian and post-Newtonian contributions along with the fourth order cumulant $\kappa_4$ in Fig.(\ref{PN}) for a BEC with $N_0\sim 10^9$ and $\omega\sim 100$ Hz. We observe that very near the cut-off time $t_{\text{min}}$ the $\kappa_4$ values goes to zero and the post-Newtonian contribution results in a reduction of the SNR when the observation time is very small. However, for large values of $t$ this behavior vanishes leading to a gain in the overall value of the cumulant $\kappa_4$ and eventually the SNR. We find out that after $t_0\sim 442.11$ sec $\kappa_4$ becomes greater than the leading order Newtonian contribution and this gain is purely due to the post-Newtonian contribution. This signifies a very important outcome and dependence on the post-Newtonian contribution which is a bit non-trivial and also the main result in our manuscript.  A more rigorous analysis requires a side by side comparison with the experimental noise-spectrum for such BEC-based model system which is out of the scope of our current manuscript.

\section{Conclusion}\label{S5}
\noindent In this work, we consider the Bose-Einstein condensate which is creating quantum fluctuation in the background which results in a skewness in the Gaussian profile of the model system. We start by analyzing the model presented in \cite{BECPRXQuantum} and extended it to the post-Newtonian setting. We start by deriving the interaction Hamiltonian in presence of the post-Newtonian background and arrive eventually at the total Hamiltonian operator of the model system. We have then obtained the time evolution equation of the raising and lowering operators. Using algebraic techniques, we have then derived a normalized form of the exponential function of the number operators which is an  important finding in the paper. Considering coherent states of the Bose-Einstein condensate, we have then obtained the expectation values of the quadrature operator and its higher orders which finally helped us to obtain the analytical expression of the fourth order cumulant. The non-vanishing fourth order cumulant indicates the existence of quantum gravitational signatures in the signal to noise ratio itself. We find out that the post-Newtonian contribution becomes significant around the minimum value of time $t=t_{\text{min}}$, at which point the SNR becomes zero. In the range $t_{\text{min}}\leq t\leq t_0$, $|\kappa_4|\leq |\kappa_4^{(0)}|$, where $|\kappa_4^{(0)}|$ is the cumulant due to purely Newtonian contribution. However, after the time $t_0$, $|\kappa_4|$ becomes greater than $|\kappa_4^{(0)}|$. This non-trivial behavior of the cumulant leading to the SNR comes only due to the post-Newtonian contributions and may lead to significant changes in the known SNR profiles of BEC based quantum gravity detectors in future. Finally, a more rigorous analysis can be conducted using squeezed coherent states investigating the effects of squeezing on such post-Newtonian terms in the signal to noise ratio.
This we leave for the future.

\section{Appendix A}
\noindent In this part we have extended our analysis to a squeezed vacuum state-$|\xi\rangle$. It is evident that a similar approach may be done in calculating the fourth order cumulant-$\kappa_4$. Using the analytical expression of $\hat{a}^n(t)$ in eq.\eqref{1.32}, one needs to calculate $\langle \hat{q}^n(t)\rangle$ in a squeezed vacuum state. Before proceeding to obtain the final expression for $\kappa_4$, we recall the following standard relations; in particular we note 
\begin{equation}
    N_0=\sinh^2{r}; \quad \xi=re^{i\phi}
\end{equation}
where r is the squeezing amplitude which determines the magnitude of squeezing \& $\phi$ is the squeezing phase angle. Here, $N_0$ is the average number of particles in the condensate. These relations will be helpful in determining the fourth order cumulant. Performing similar calculations as in the previous section(s) and  keeping terms up to $\mathcal{O}(\chi_{PN})$, we can then obtain the leading order analytical expression for the fourth cumulant $\kappa_4$ as
\begin{equation}
    \begin{split}
    |\kappa_4|&\approx\chi_NtN^3_0\left(11\sin{2\nu}+24\sin{\nu}+10\cos{\nu}\right)\\&+\chi_{PN}tN^4_0\left(10\cos{\nu}+40\sin{\nu}\right) 
  \end{split}  
\end{equation}
where $\nu\equiv\phi-2\theta$. We can now set the phase $\nu$ to $\frac{\pi}{2}$ to obtain the leading order expression of $\kappa_4$ in this post-Newtonian analysis as 
\begin{equation}\label{5.59}
    |\kappa_4|\approx24\chi_NtN^3_0+40\chi_{PN}tN^4_0
\end{equation}
This scaling is in agreement with Ref\cite{PhysRevA.87.033839} at $\chi_N^4$ with $\theta=\pi/2$ at the Newtonian order. Since in our model leading base line remains Gaussian and hence in the denominator only the leading Gaussian contribution actually contributes to the signal to noise ratio. Thus, for calculating the denominator of the expression eq.\eqref{1.26}, we use the previously used form of $var(k_4)$, where for a large number of independent measurements i.e.$(\mathcal{M}\gg1)$,eq.(\ref{1.27}) can be approximated to
\begin{equation}\label{5.60}
\begin{split}
\text{var}(k_4)&\approx \frac{1}{\mathcal{M}}\Bigg[\kappa_8+16\kappa_2\kappa_6+48\kappa_3\kappa_5+34\kappa_4^2+72\kappa_2^2\kappa_4\\&+144\kappa_2\kappa_3^2+24\kappa_2^4\Bigg]
\end{split}
\end{equation}
However, this is an approximate relation and is limited to the regime used in this section. Calculating the required cumulant values individually for a squeezed state in eq.(\ref{5.60}) one can get the analytical expression of $\sqrt{var(k_4)}$ as
\begin{equation}\label{5.61}
    \sqrt{var(k_4)}\approx\frac{4.9(\cosh{2r}-\cos{\nu}\sinh{2r})^2}{\sqrt{\mathcal{M}}}
\end{equation}
The expression for SNR using eq.(s)(\ref{1.26},\ref{5.59},\ref{5.61}) can then be finally found to be
\begin{equation}
    SNR=\frac{4.9\chi_NtN^3_0+8.2\chi_{PN}tN^4_0}{(\cosh{2r}-\cos{\nu}\sinh{2r})^2}\sqrt{\mathcal{M}}
\end{equation}
Here, we may note that for large $N_0\gg1$, one may approximate the SNR expression to be 
\begin{equation}
    SNR\approx4.9\left(\chi_NtN_0\sqrt{\mathcal{M}}+1.7\chi_{PN}tN^2_0\sqrt{\mathcal{M}}\right)
\end{equation}
However, when $\chi_{PN}$ is not so small, the above SNR approximation may not be accurate and in such cases one may look at the SNR for squeezed coherent states.

\end{document}